\documentclass[twocolumn,amsmath,amssymb, superscriptaddress]{revtex4}

\usepackage{graphicx}
\usepackage{dcolumn}
\usepackage{bm}
\usepackage{epstopdf}
\usepackage{color}

\begin{document}


\title{Characterization of periodic cavitation in an optical tweezer}

\author{Viridiana Carmona-Sosa}%
\affiliation{Instituto de Ciencias Nucleares, Universidad Nacional Aut\'onoma de M\'exico\\ Apartado Postal 70-543, 04510, M\'exico D.F., M\'exico.}
\author{Jos\'e Ernesto Alba-Arroyo}
\affiliation{Instituto de F\'isica, Universidad Nacional Aut\'onoma de M\'exico\\ Apartado Postal 20-364, 01000, M\'exico D.F., M\'exico.}
\author{Pedro A. Quinto-Su}%
\email[E-mail: ]{pedro.quinto@nucleares.unam.mx}
\affiliation{Instituto de Ciencias Nucleares, Universidad Nacional Aut\'onoma de M\'exico\\ Apartado Postal 70-543, 04510, M\'exico D.F., M\'exico.}


\begin{abstract}
Microscopic vapor explosions or cavitation bubbles can be generated periodically in an optical tweezer with a microparticle that partially absorbs at the trapping laser wavelength. 
In this work we measure the size distribution and the production rate of cavitation bubbles for microparticles with a diameter of 3 $\mu$m using high speed video recording and a fast photodiode. We find that there is a lower bound for the maximum bubble radius $R_{max}\sim 2~\mu$m which can be explained in terms of the microparticle size.
More than $94 \%$ of the measured $R_{max}$ are in the range between 2 and 6 $\mu$m, while the same percentage of the measured individual frequencies $f_i$ or production rates are between 10 and 200 Hz.
The photodiode signal yields an upper bound for the lifetime of the bubbles, which is at most twice the value predicted by the Rayleigh equation. We also report empirical relations between $R_{max}$, $f_i$ and the bubble lifetimes.
\end{abstract}

\maketitle


\section{Introduction}
Laser-induced cavitation involves the fast deposition of laser energy in a liquid, resulting in an explosion that creates a transient vapor bubble that expands reaching a maximum radius $R_{max}$ and then collapses under the static pressure of the surrounding liquid \cite{tomita}. These bubbles have been used in many different applications where fast actuation of liquid or where impulsive forces are required at a microscopic length scale. For example, in high speed microfluidics \cite{highspeed1} and micro-pumps \cite{micropump}, cell membrane permeabilization \cite{nbubble5, nbubble1} and cell lysis \cite{lysis1},  red blood cell stretching and poration \cite{rbcp,rbc1},  malaria detection \cite{malaria1}, bending of carbon nanotubes and nanowires \cite{cnt1,nw1}. 

In the last few years there has been a lot of interest in the production of cavitation bubbles through photothermal processes in nanomaterials \cite{nbubble5,nbubble1,nbubble2,nbubble3,nbubble4} resulting in bubbles with sizes on the order of several hundreds of nanometers. These transient nanobubbles have shown great potential for medical treatments \cite{nano1}. Optical tweezers or highly focused continuous (CW) lasers have been used to create micron sized cavitation bubbles in liquids that are transparent to the laser wavelength \cite{dholakia1, lajoinie1}. Dholakia and coworkers \cite{dholakia1} trapped a polystyrene nanoparticle that was later irradiated with a nanosecond laser pulse creating optical breakdown at the trapped nanoparticle. 
Another approach used custom made microparticles filled with absorbing dye that was rapidly vaporized creating multiple cycles of bubble expansion and collapse in a timescale of a few microseconds \cite{lajoinie1}.  

Recently we showed that a microparticle that partially absorbs at the wavelength of the trapping beam in an optical tweezer can periodically generate microscopic explosions \cite{steam1}. 
The particle is simultaneously heated and pulled towards the waist of the trapping beam where it superheats the liquid and creates a transient vapor explosion that pushes the particle below the waist, where the cycle restarts (Fig. 1a). 
However, in that study we could not establish the bubble size distribution of $R_{max}$ and production rate, which are important for potential applications. In the present work we measure hundreds of bubbles with two different methods: fast video recording and a fast photodiode which dramatically increases the time resolution. We find that there is a minimum bubble size of $R_{max} \sim ~2\mu$m and that most of the bubbles have an $R_{max}$ between 2 and 6$~\mu$m which is a fairly narrow range.

\begin{figure*}
\includegraphics[width=5.4in]{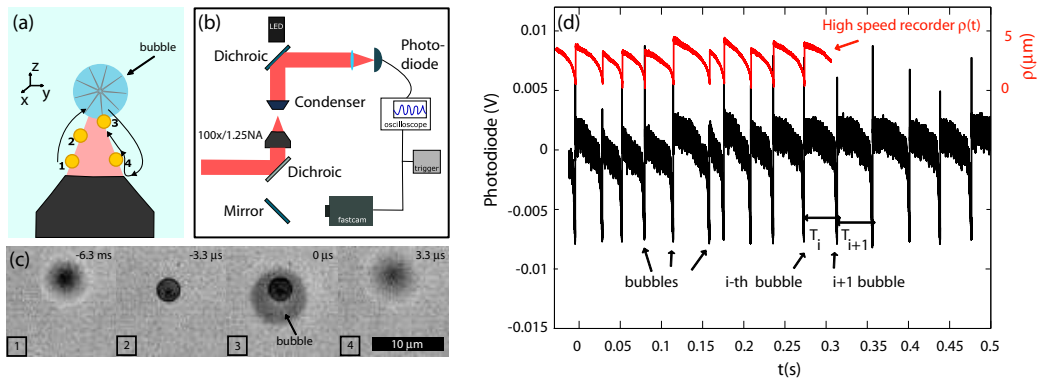} 
\caption{(a) Periodic cavitation in an optical tweezer. (b) Experimental setup. (c) Selected frames corresponding to (a) from a high speed recording at 300,000 fps. (d) Light scattered by the microbead and the bubble detected with the photodiode (black line). Particle position $\rho (t)$ in the $XY$ plane (red line), $\rho (t) =\sqrt{(X(t)-X_0)^2+(Y(t)-Y_0)^2}$ where $X_0$, $Y_0$ are the coordinates of the beam center.}
\label{exp} 
\end{figure*}

\section{Experiment}
The experimental setup is a near IR optical tweezer with a wavelength of 975 nm  focused by a 100X/1.25 NA microscope objective. The microparticles are standard magnetic beads (Bangs labs) with a mean diameter of 3.16 $\mu$m immersed in water. In the experiments, the microparticles interact individually with the focused laser. The liquid sample is placed between two No. 1 microscope glass coverslips (thickness $0.13-0.16~$mm) separated with spacers with a height $\sim 100 ~\mu$m. The sample is then sealed with nail polish. Some experiments were done without sealing the sample and the results were consistent with those with the sealed chamber. 
A schematic of the experimental setup is depicted in Fig. 1b. The transmitted laser power by the microscope objective is between 50 and 62 mW. The trapping laser light is collected with the microscope condenser and a lens into a fast photodiode (175ps rise, Alphalas) which replaces the quadrant detector in a typical optical tweezer setup.  The photodiode is connected to an oscilloscope (1GHz, LeCroy 610Zi). The sample is illuminated by an LED and imaged into a high speed video recorder (Photron SA 1.1). 

The oscilloscope and the high speed video recorder are triggered with a delay generator (SRS DG535). The high speed video recorder captures between 100,000 and 150,000 frames at 300,000 frames per second (fps). The oscilloscope records 25 million points in a time span of one second which yields a time step $\Delta t$ of 40 ns, sufficient to measure the width of the signal of the scattered light during cavitation.
Typical images extracted from a high speed video recording of a single cycle are shown in Fig. 1c. First the particle is below the focus of the microscope objective which is why it is blurred, then it is pulled towards the geometrical focus where the particle appears sharper. The cavitation bubble appears in one frame and in the next ($3.3~\mu$s later) the bubble has vanished and the particle is below the focus of the objective.  

A sample of an oscilloscope trace for the scattered light collected by the photodiode is shown in Fig. 1d (black line, 0-0.5 s). The bubbles appear at the sharp transitions, labeled as the i-th bubble which is followed by a time interval $T_i$ (cycle period) that depends on the size of the bubble. The frequency of that cycle is $f_i=1/T_i$. The red line corresponds to the distance of the microparticle in the $XY$ plane to the center of the trap (located at $(X_0, Y_0)$) where the laser beam is focused.

\begin{figure*}
\includegraphics[width=5.4in]{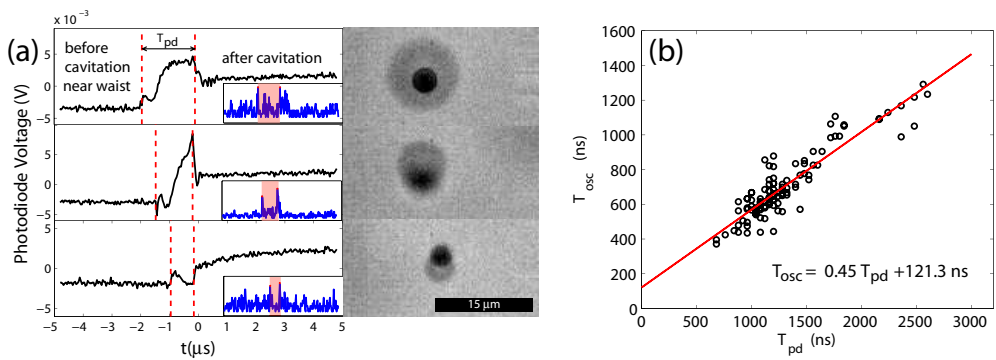}
\caption{(a) Left column: Photodiode signals during cavitation events for different bubble sizes. Inset: Absolute value of the numerical first derivative in the same temporal range. (a) Right column: Frame where the bubble corresponding to the event in the left appears. 
(b) Data for calculated spherical oscillation time (from measured $R_{max}$) as a function of the width of the photodiode signal (116 events).
}
\end{figure*}

\section{Results}
Figure 2a (Left) shows the zoomed photodiode signal for selected cavitation events and the corresponding images for the bubble and particle extracted from the high speed videos. We observe that there is a change in the voltage level which corresponds to the particle being close to the waist of the beam (before cavitation) and after the particle has been pushed by the bubble below the waist. The transition signal in between captures the large particle displacement and the bubble dynamics which is why it can have different shapes (peak or a valley) depending on the spatial location where cavitation occurs and the direction in which the particle is pushed due to rapid changes in Mie scattering by both particle and bubble. Usually the transition starts with a small peak followed by a larger one, hence it is not straighforward to locate the part that corresponds to the bubble. 
The width $T_{pd}$ of the transition is defined by numerically differentiating (finite difference) the signal, taking the absolute value and measuring the time between the two largest peaks (insets in Fig. 2a Left). 
One more condition is that the two maximums (in the derivative) have to span the whole transition (including the smaller first peak), in order to get the maximum width and hence an upper bound to the bubble lifetime $T_{pd}$ measured with the photodiode. This criteria yields 679 events. 

We measure the maximum bubble radius $R_{max}$ from a single frame in the high speed videos where the bubble appears blurred (Fig. 2a Right). The error in the measurement of $R_{max}$ is $\pm 0.22 ~\mu$m estimated by varying the graylevel thresholds that define the edge of the bubble. We were able to measure 179 bubbles where the image had good signal to noise ratio. 
To estimate the bubble lifetime (the time it takes the bubble to fully expand and collapse) we assume spherical symmetric bubble dynamics  \cite{vogel, nbubble3}. The spherical bubble lifetime is $T_{osc}$ which is twice the collapse time $T_C$ that can be calculated using the Rayleigh formula \cite{rayleigh1907}: 
\begin{equation}
T_{osc}=1.82 R_{max} \sqrt{ \frac{\rho _l}{p_0 -p_v} }
\end{equation}
where $\rho _l $ is the density of the liquid (water), $p_0$ the atmospheric pressure (0.1 MPa) and $p_v$ the vapor pressure at room temperature (2330 Pa at 20$^\circ$C ). 
However there is a loss in the accuracy of the Rayleigh formula for  nanobubbles \cite{nbubble2} (since it does not take into account surface tension). In this study the maximum deviation should be at most a factor of 1.2 \cite{vogel1} for the smallest bubbles $R_{max}<2.7~\mu$m. Hence we use $T_{osc}$ to get a lower bound for the bubble lifetime. The spherical shape assumption is a strong one but it has been used successfully to model microparticles accelerated by cavitation bubbles created at the microparticle surface \cite{partaccel}. Furthermore, some experimental measurements on the dynamics of photothermal bubbles show symmetric dynamics \cite{nbubble1}.

For some events (116) we were able to measure simultaneously $T_{pd}$ and $R_{max}$. The plot of the calculated oscillation time $T_{osc}$ (extracted from the measured $R_{max}$) as a function of the width of the photodiode signal $T_{pd}$ is in Fig. 2b. The bubble lifetime predicted by the Rayleigh formula is proportional to the measured width of the photodiode signal during the transition: $T_{osc}= 0.45 T_{pd} +121.3$ ns.
The width of $T_{pd}$ is an upper bound to the lifetime, so Fig. 2b shows that the lifetime is at most two times that predicted by Rayleigh equation for $T_{osc}$ [Eq. (1)]. Deviations could come from non spherical dynamics, compressibility and other photothermal processes that have been neglected.

\begin{figure*}
\includegraphics[width=5.4in]{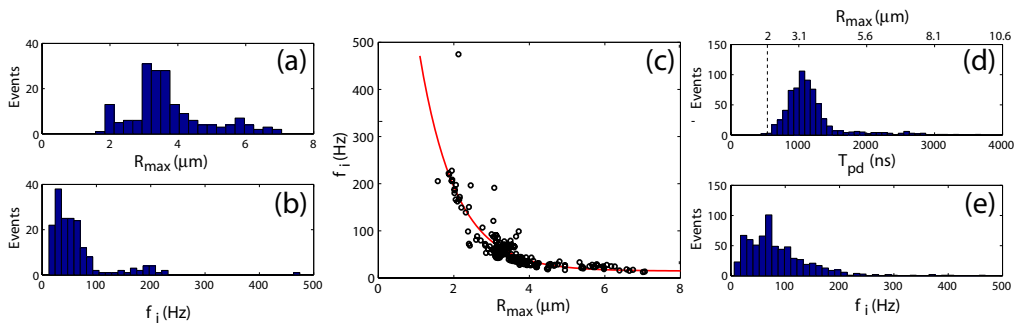}
\caption{ (a) Histogram for measured $R_{max}$ (video data 179 events). (b) Histogram for measured $f_i$ (video data). (c) Cycle frequency $f_i$ as a function of maximum bubble radius $R_{max}$ (video data).
(d) Histogram for measured $T_{pd}$ (photodiode data 679 events). (e) Histogram for measured $f_i$ (photodiode data).
}
\end{figure*}

Figure 3(a-c) depicts the data extracted from the high speed videos (179 events).
The histograms in in Figs. 3a and 3b correspond to the measured $R_{max}$ and the individual frequencies $f_i$ respectively. The largest bars are: 31 events where $R_{max}$ has values between $2.95-3.22~\mu$m and 38 events where $f_i$ has a value between $24.1-35.7$ Hz. There are 13 events with $R_{max}$ between 1.85-2.12$~\mu$m,  one event with an $R_{max}$ of 1.6$~\mu$m and no events with a smaller $R_{max}$.
The individual cycle frequencies $f_i$ are plotted in Fig. 3c as a function of the measured $R_{max}$. The red solid line is a fit to an exponential decaying function $f_i= a+b \exp{(-c R_{max})} $, where $a= 14.7~$Hz, $b=1434.8~$Hz and $c=1.0412~\mu\mathrm{m}^{-1}$. The fit is reasonable up to frequencies of about 200 Hz. This could be useful to estimate bubble size by just measuring the frequency.

The photodiode data (679 events) are shown in Figs. 3d and 3e which are histograms for $T_{pd}$ and $f_i$ respectively. The largest bars in the histograms are 106 events with $T_{pd}$ between 1010-1090 ns and 106 with $f_i$ between 63.9-75.5 Hz. 
The data in Fig. 3d can be used to get $R_{max}$ by first using the linear relation between $T_{osc}$ and $T_{pd}$ and then solving for $R_{max}$ in [Eq. (1)]. There are 106 events with $R_{max}$ between $3.15-3.34 ~\mu$m (Fig. 3d).
Surprisingly we get the same lower bound for $R_{max} \sim 2~\mu$m as in Fig. 3a, considering that the time resolution with the photodiode is much greater than with the video camera and can resolve very short $T_{pd}$. There is 1 event corresponding to $1.85~\mu$m, 3 events at $1.95-2.15~\mu$m and 17 at $2.15-2.35~\mu$m.   

\section{Discussion and Analysis}
The parameters that describe the interaction between a spherical microparticle and a cavitation bubble are the microbead radius $R_p$, the maximum bubble radius $R_{max}$, the initial separation between the microparticle and the origin of the bubble $\delta '$ and the displacement of the particle $\Delta x '$ \cite{transientflow, nw1}. The interaction is described by a mastercurve for $\Delta x$ as a function of $\delta x$ where $\delta =\delta '/ R_{max}$ and $\Delta x = \Delta x '/(2R_p)$. 
After the bubble collapses the microparticle can stop farther away from the starting point (repulsion $\Delta x>0$ for $\delta<0.7$), closer to the starting position (attraction $\Delta x <0$ for $0.7<\delta <3.7 $) or at the same distance (neutral $\Delta x =0$ for $\delta =0.7$). 
 
During the bubble expansion the particle is displaced in a timescale of microseconds (repulsion $\Delta x >0$, $\delta<0.7$), while the return to the beam waist with optical forces is much slower in the tens of milliseconds. 
Hence it is expected that the bubbles with the smaller $R_{max}$ will induce a smaller displacement and a shorter time back to the position of the next vapor explosion. 
Since the bubble starts at the surface of the microparticle then $\delta ' = R_p=1.58 ~\mu$m. The equilibrium position for $\delta = 0.7 $ would correspond to the maximum frequency since essentially the particle would end at the same spot after the cavitation bubble collapses where it should be superheated again in a very short time producing another bubble. For $\delta =0.7$, $R_{max}=2.26 ~\mu$m, which is consistent with the smallest measured $R_{max}$ (Figs. 3a, 3d) which correspond to the largest frequencies (Fig. 3c). 

The width in the histograms for $f_i$ presented in Figs. 3b and 3e is larger than that for $R_{max}$ and $T_{pd}$. This can be explained considering the random direction in which the particle is pushed which results in different optical forces ($\propto \nabla I$, with $I$ the intensity) and hence different speeds. 

The distribution of $R_{max}$ in Figs. 3a and 3d is consistent with the previous measurements \cite{steam1} where only 40 bubbles were measured and smaller bubbles were not detected due to insufficient signal to noise caused by poorer illumination of the sample. 
Here the measured frequencies lie in a larger interval than that reported in \cite{steam1} (23 to 48 Hz) because in that study the frequencies were extracted from the largest peaks on the Fourier spectrum of the microparticle dynamics. 

\section{Conclusion}
We have measured distribution of $R_{max}$ which is surprisingly narrow with most values between 2-6$~\mu$m. The size is comparable with bubbles created with focused femtosecond laser pulses \cite{vogel1} and some photothermal bubbles in nanomaterials \cite{nbubble1}. The measured cutoff or lower bound for the bubble sizes of $\sim 2~\mu$m agrees with the neutral displacement regime reported in \cite{transientflow}. We also found that the production rate of the cavitation bubbles depends in the bubble size as expected, since a larger bubble results in a larger displacement for the microparticle and a slower return back to the waist of the trapping beam. 
The width of the photodiode signal yielded upper bounds to the lifetime of the bubbles which are at most twice those predicted by the Rayleigh formula for $T_{osc}$. 
The empirical relations between $R_{max}$, $T_{osc}$, $T_{pd}$ and $f_i$ could be useful to estimate $R_{max}$ and the lifetime when a fast video recorder is not available.  
Further studies in the shape and dynamics of the bubbles could include ultra-high speed imaging with temporal resolution of tens of nanoseconds in order to measure the full dynamics and shape of the bubbles.

Work partially funded by DGAPA-UNAM project IN104415 and CONACYT National Laboratory project LN260704.

\begin{acknowledgments}
Work partially funded by DGAPA-UNAM project IN104415 and CONACYT National Laboratory project LN260704. 
\end{acknowledgments}


\end{document}